\documentclass[twocolumn]{biophys}

\usepackage{helvet,times}
\usepackage{booktabs}

\usepackage{array}
\usepackage{multirow}
\usepackage{tabularx}

\usepackage{amsmath}
\usepackage{amssymb,amsfonts,amsthm,bm}
\usepackage[round,numbers,sort&compress]{natbib}
\usepackage[utf8]{inputenc}

\usepackage[hidelinks]{hyperref}
\usepackage{courier}

\jno{kxl014} 
\gridframe{N} 
\cropmark{N} 

\doi{doi: 10.1529/biophysj.XYZ.123456} 

\begin{document}

\setcounter{page}{1} 
\title{GIBS: A grand-canonical Monte Carlo simulation program for simulating ion-biomolecule interactions}

\author{Dennis~G.~Thomas\thanks{Biological Sciences Division, Pacific Northwest National Laboratory}, Nathan~ A.~Baker\thanks{Advanced Computing, Mathematics, and Data Division, Pacific Northwest National Laboratory; Division of Applied Mathematics, Brown University}}

\begin{abstract}
{The ionic environment of biomolecules strongly influences their structure, conformational stability, and inter-molecular interactions. This paper introduces GIBS, a grand-canonical Monte Carlo (GCMC) simulation program for computing the thermodynamic properties of ion solutions and their distributions around biomolecules. This software implements algorithms that automate the excess chemical potential calculations for a given target salt concentration. GIBS uses a cavity-bias algorithm to achieve high sampling acceptance rates for inserting ions and solvent hard spheres in simulating dense ionic systems. In the current version, ion-ion interactions are described using Coulomb, hard-sphere, or Lennard-Jones (L-J) potentials; solvent-ion interactions are described using hard-sphere, L-J and attractive square-well potentials; and, solvent-solvent interactions are described using hard-sphere repulsions. This paper and the software package includes examples of using GIBS to compute the ion excess chemical potentials and mean activity coefficients of sodium chloride as well as to compute the cylindrical radial distribution functions of monovalent (Na$^+$, Rb$^+$), divalent (Sr$^{2+}$), and trivalent (CoHex$^{3+}$) around fixed all-atom models of 25 base-pair nucleic acid duplexes. GIBS is written in C++ and is freely available for community use; it can be downloaded at \url{https://github.com/Electrostatics/GIBS}.}


{Submitted to Biophysical Journal (Article Type: Computational Tools)}

{Please address correspondence to Dennis Thomas (dennis.thomas@pnnl.gov) and Nathan Baker (nathan.baker@pnnl.gov)}
\end{abstract}

\maketitle 

\section*{Introduction}
Biomolecules, such as nucleic acids, are highly charged systems.
Therefore, their structure, conformational stability, and molecular interactions are determined by their molecular composition as well as their ionic and solvent environments \cite{lipfert2014understanding}.
A topic that continues to interest the biophysical research community is how monovalent and multivalent counterions govern a wide range of biological processes such as genome packaging (in cellular compartments and viral capsids) \cite{nguyen2013strongly, knobler2009physical, fuller2007ionic, maffeo2014close, allahverdi2015chromatin}, RNA folding \cite{misra1998role, draper2004guide, draper2005ions, mak2012ions}, ribosome activity \cite{fedor2009comparative}, protein-nucleic acid interactions \cite{von2007simple, stickle2007cation, mackerell2008molecular}, ligand binding \cite{misra1994salt, sharp1995salt}, and transitions between left-handed and right-handed forms of DNA and RNA duplexes \cite{pan2014ion}.
Understanding these processes through computational modeling requires accurate and efficient models for the ionic environment \cite{lipfert2014understanding}, which has been the goal of various
theoretical studies (e.g., counterion condensation  \cite{manning1969limiting} and Poisson-Boltzmann (PB) theories \cite{anderson1982polyelectrolyte}) and computational modeling studies based on the PB equation \cite{anderson1982polyelectrolyte}, molecular dynamics simulations \cite{maffeo2014close, pan2014ion, rueda2004exploring, varnai2004dna, burak2004competition, allahyarov2003adsorption, yoo2012competitive, korolev2003molecular, korolev2002competition, korolev2001spermine, pabit2016understanding,
drozdetski2016opposing, tolokh2016multi, tolokh2014double}, Monte Carlo simulations (in canonical and grand canonical ensembles) \cite{olmsted1995grand, lyubartsev1997monte, korolev2001competitive, korolev2001spermine, korolev1999competitive,nguyen2016grand}, and classical density functional theories \cite{sushko2016role,ovanesyan2014excluded}.

All-atom MD simulations provide the most detailed approaches for predicting the structure and dynamics of ion atmosphere and their effects on solute conformations; however, they are computationally time-consuming.
Alternatively, much can be learned about the structure and thermodynamic properties of ionic solutions at equilibrium, using coarse-grained models of the solvent (and molecular ions) and effective ion-ion potentials based on MD-derived potentials of mean force (PMFs).
The development of such coarse-grained models involves extensive parameterization and testing against the thermodynamic properties of electrolyte solutions (e.g., mean-activity coefficients), ion counts and distributions around biomolecules at equilibrium.
Experimental data on salt activity, ion counts and scattering data, can then be used to check the accuracy and domain of applicability of these models with different simulation methodologies. 

Among the various simulation methodologies, grand-canonical Monte Carlo (GCMC) methods have been popular for predicting the thermodynamic properties of ion solutions and their distributions around biomolecules \cite{vitalis2004isim,vincze2010nonmonotonic,
olmsted1995grand,malasics2010efficient,lamperski2007individual,
valleau1980primitive,zhang1993simulations,jayaram1991grand,
wu1999grand,vlachy1986grand,jayaram1996modeling,
lenart2007effective,korolev1999competitive,
korolev2001spermine,korolev2001competitive,nguyen2016grand}.
The GIBS software program uses GCMC methods for two tasks.
Unlike our previous GCMC software program, ISIM \cite{vitalis2004isim}, GIBS was developed with three specific objectives in mind: to automate excess chemical potential calculations for bulk solutions; to enable fast and efficient GCMC sampling of ion distributions across a range of concentrations; and to serve as a GCMC simulation platform for testing new coarse-grained ion and solvent models.

\section*{Methodology}

\subsection*{Sampling methods}
The current version of GIBS treats the ions and solvent species as spherical particles with a formal charge (as applicable) at their centers.
The simulation box is a rectangular cuboid of volume $V = L_{x} \times L_{y} \times L_{z}$; where $L_{x}$, $L_{y}$, and $L_{z}$ are the box lengths along the three Cartesian coordinate axes.
GIBS uses three standard types of GCMC moves to sample the distribution of ion/solvent species: insertion, deletion, and displacement.

In the insertion step, a particle is placed at a random location inside the simulation box; in the deletion step, a randomly selected particle species is removed; and, in the displacement step, a particle is randomly selected and displaced. Each move is accepted with a probability (see formulas in Supporting Material). GIBS uses the cavity-biased grid-insertion algorithm \cite{woo2004grand} to achieve high sampling efficiency when simulating dense systems containing ions and solvent hard spheres. The algorithm identifies the cavities available for inserting each particle type on a cubic lattice grid. It then randomly selects a cavity and places the particle at the center of the cavity.

\subsection*{Calibration of ion excess chemical potentials for bulk electrolyte solutions}
The excess chemical potential is an important parameter for achieving desired bulk concentrations for new models of ion and solvent species.
GIBS implements two methods to calibrate the excess chemical potential, $\mu_{i}^{\text{ex}}$, to achieve the bulk concentration of the particle species $c_{i}$ for the particular species model: the iterative charge-corrected adaptive GCMC algorithm (A-GCMC) \cite{malasics2010efficient} and the proportional-integral-derivative (PID) control approach \cite{speidel2006automatic}. A brief description of the two methods is also provided in the Supporting Material.
The corresponding (configurational) chemical potential for particle species $i$, is computed using the equation, 
\begin{equation}
	\mu_{i} =  k_{B}T \ln c_{i} + \mu_{i}^{\text{ex}}.
\end{equation}
Since the number of each ion species fluctuates at each step of the GCMC simulation, the total charge in the system also fluctuates at each step: the system is only electrically neutral on average. The excess chemical potential is used to compute the mean activity coefficients of salts for a given bulk electrolyte concentration.

\subsection*{Ion and solvent interaction models}
Three types of solvent-ion interaction models have been implemented in GIBS.
The simplest model is the unrestricted primitive model (UPM) which represents ions as charged hard spheres and solvent (water) as a dielectric continuum.
Two variations of the solvent primitive model (SPM) are also implemented:  a simple version where the solvent is represented as neutral hard spheres to account for excluded solvent effects and a more detailed version that includes Lennard-Jones ion-water interactions (SPM+LJ$_\text{IW}$).
Long-range electrostatic interactions between charged hard sphere ions are computed using the pair-wise Coulomb potential in a uniform dielectric (equal to the solvent dielectric permittivity).
Periodic boundary conditions with the minimum image convention \cite{frenkel1996understanding} are used for all non-bonded interactions.

\subsection*{Ion/solvent-solute interactions}
The interaction energy between a fixed solute (e.g., nucleic acid) and each ion charge $q_i$ in the simulation box is explicitly calculated as 
\begin{equation} \label{eq:ion_solute}
	U_{i}(r_i) =  q_i \phi(r_i) + u_i(r_i)
\end{equation}
where $\phi_{i}$ is the solute electrostatic potential at location $r_i$ of the ion charge and $u_i$ is the steric potential experienced by an ion of species $i$ at location $r_i$.
The potential $\phi$ is determined by solving Poisson's equation with the Adaptive Poisson Boltzmann Solver (APBS) software package \cite{baker2001electrostatics} (\url{http://www.poissonboltzmann.org}).
This approximation, along with a fixed solute conformation is preferred because it requires the Poisson equation to be solved only once (at the beginning) for the entire GCMC simulation.

The ion/solvent-solute steric interactions $u_i$ are described by a hard sphere potential between each particle (ion/solvent) type and a solute atom.
Specifically, the steric potential is infinite for atoms closer than the sum of their radii and zero otherwise.
In the grid-insertion algorithm, the non-accessible region for each particle type due to the solute atoms is determined by identifying the cavity grid cells that fall within the closest distance of approach, before the start of the simulation.
Eq.~\ref{eq:ion_solute} neglects the desolvation potential induced by bringing a charge close to the low-dielectric interior of biomolecular solutes as well as ion-biomolecule Lennard-Jones interactions  \cite{bostrom2005energy, duignan2013continuum}.

\subsection*{Software compilation and use}
The GIBS software is written in C++ and can be downloaded from \url{https://github.com/Electrostatics/GIBS}.
The source code is compiled using the Makefile that is generated using CMake \cite{martin5mastering}.
Instructions on how to build and compile the code are provided with the software package. 
To run a simulation, the user needs to create a simulation directory, and two sub-directories:  \texttt{inputfiles} and  \texttt{outputfiles}.
The input files are set up in the \texttt{inputfiles} folder and the output files from the simulation are generated in the \texttt{outputfiles} folder.
Example simulation files and results (for testing) are also provided with the software.

\section*{Example Applications}
Four example applications are provided with the GIBS software, as described below.
In all of the examples, we used Pauling radii for the Na$^{+}$ (1.02~\r{A}), Rb$^{+}$ (1.49~\r{A}), and Sr$^{2+}$ (1.25~\r{A}), and Cl$^{-}$ (1.81~\r{A}) ions.
The CoHex$^{3+}$ radius was set to 3~\r{A}, the value used by Tolokh et al \cite{tolokh2014double} in continuum electrostatics calculations.
A uniform dielectric constant of 78.5 and temperature of 298 K was used in all simulations. The grid spacing for insertion cavity detection was set to 0.5 \r{A}. The solute dielectric constant for the duplexes in the Poisson calculations was set to 2.
The LJ potential parameters used in the $\text{SPM+LJ}_\text{IW}$ simulations are shown in Table~\ref{tab:lj_parameters}. All the simulations were run on a 64-bit Windows 7 Operating System with 16 GB RAM and two 1.60 GHz Intel Xeon\textsuperscript{\textregistered} 6-Core processors using MINGW C++ compiler.
\begin{table}
	\centering
	\caption{\label{tab:lj_parameters}Lennard-Jones parameters.}
	\begin{tabular}{|>{\centering\arraybackslash}m{1 cm}|>{\centering\arraybackslash}>{\centering\arraybackslash}m{2cm}|>{\centering\arraybackslash}m{1.5cm}|>{\centering\arraybackslash}m{1.8cm}|}
		\hline
		Particle type & L-J collision diameter, $\sigma$ (\r{A}) & L-J well depth, $\epsilon$ (kcal/mol) & Reference\\
		\hline
		Na$^+$ & 2.159538 & 0.352642 & \cite{joung2008determination} \\
		\hline
		Rb$^+$ & 3.094982 & 0.445104 &  \cite{joung2008determination} \\
		\hline
		Sr$^{2+}$ & 3.100000 & 0.059751 & \cite{mamatkulov2013force} \\
		\hline
		Cl$^-$ & 4.400000 & 0.100000 &  \cite{mamatkulov2013force} \\
		\hline
		Water & 3.166000 & 0.155354 & SPC/E  \cite{berendsen1987missing} \\
		\hline
	\end{tabular}
\end{table}
In the SPM and $\text{SPM+LJ}_\text{IW}$ simulations, the solvent (water) radius was set to 1.4 \r{A}, a typical value for simple water models \cite{lee1971interpretation}. 
The solvent (water) packing fraction was set to 0.3, which corresponds to a solvent concentration of 22.693 M. 
Although this concentration is low for water under standard conditions (55.5 M, $\eta = 0.73$), it has been shown to capture the steric effects imposed by the solvent hard spheres \cite{lamperski2007monte}.
Higher values of $\eta$ were not used due to the problem of non-ergodicity for simulations of hard sphere fluids at high density \cite{lamperski2007monte}. 

\subsection*{Example 1: Calculating the excess chemical potential of a NaCl solution}
\begin{figure}
	\centering
	\includegraphics{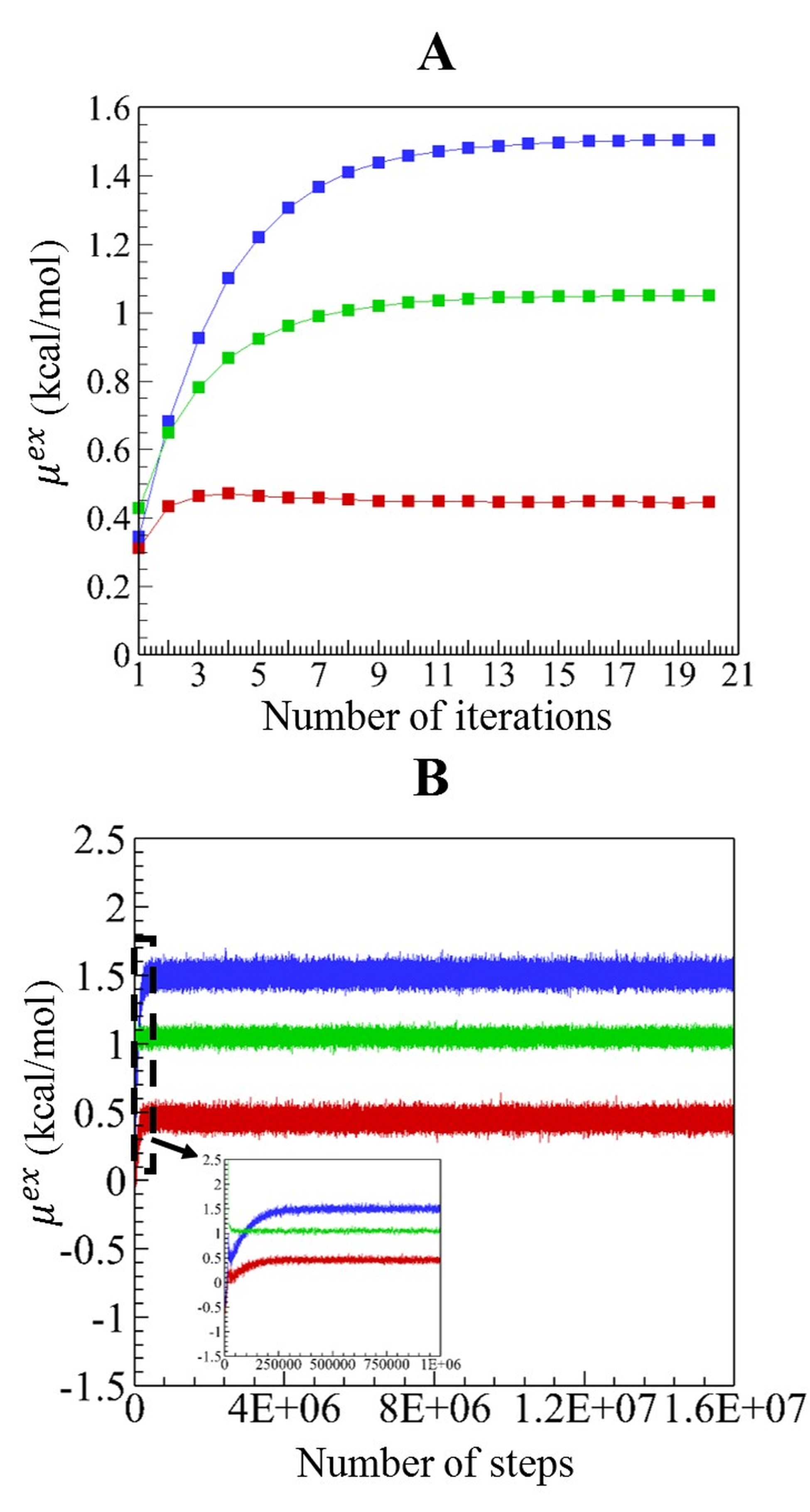}
	\caption{Convergence of the excess chemical potential ($\mu^{\text{ex}}$) for $\text{Na}^{+}$ (red), $\text{Cl}^{-}$ (blue), and water (green) in 100 mM NaCl and 22.693 M water.
	(A) The A-GCMC method converged to excess chemical potentials of $\mu_{\text{Na}}^{\text{ex}} = 0.447 \pm 0.001$, $\mu_{\text{Cl}}^{\text{ex}} = 1.503 \pm 0.003$, and $\mu_{\text{wat}}^{\text{ex}} = 1.050 \pm 0.001$ kcal/mol, and to molecule numbers of $N_{\text{Na}} = 60.1 \pm 4.5$, $N_{\text{Cl}} = 60.1 \pm 4.5$, and $N_{\text{wat}} = 13650 \pm 52.2$.
	(B) The PID method converged to $\mu_{\text{Na}}^{\text{ex}} = 0.448 \pm 0.037$, $\mu_{\text{Cl}}^{\text{ex}} = 1.509 \pm 0.041$, and $\mu_{\text{wat}}^{\text{ex}} = 1.052 \pm 0.028$ kcal/mol and $N_{\text{Na}} = 60.1 \pm 3.0$, $N_{\text{Cl}} = 60.1 \pm 3.0$, and $N_{\text{wat}} = 13665.8 \pm 13.7$.}
	\label{fig:convergence}
\end{figure}
This example compares the A-GCMC and PID methods for calculating the excess chemical potential of individual ion and solvent species in bulk electrolyte solutions using the SPM model.
The electrolyte system selected for this example was 100 mM NaCl and simulations for the excess chemical potential calculations were performed in a cubic box with 100 \r{A} side lengths. 
Each iteration in A-GCMC consisted of 10$^6$ GCMC simulation steps and $3\times10^5$ equilibration steps, and each step consisted of 7 random insertion/deletion cycles and 3 random single-particle displacement moves. 
In PID, there were $1.6 \times 10^{7}$ steps (with the same number of insertion/deletion cycles and displacement moves per step) and the values started to converge within the first $5 \times 10^6$ steps. 

The A-GCMC simulation consisted of 20 iterations, which took about 28 CPU hours to complete.
The PID simulation completed in 21 hours.

Figure~\ref{fig:convergence} shows how the excess chemical potential value converged using the A-GCMC and PID methods and a SPM NaCl/solvent model.
Although both methods gave similar results, the PID method converged much more quickly for these dense hard sphere systems. As expected, the PID simulations are noisier because the excess chemical potential is updated after every insertion/deletion step, unlike the A-GCMC method, where the updates are made only after each GCMC iteration.

\subsection*{Example 2: Calculating mean activity coefficients for various concentrations of NaCl in solutions} 
\begin{figure}
	\includegraphics{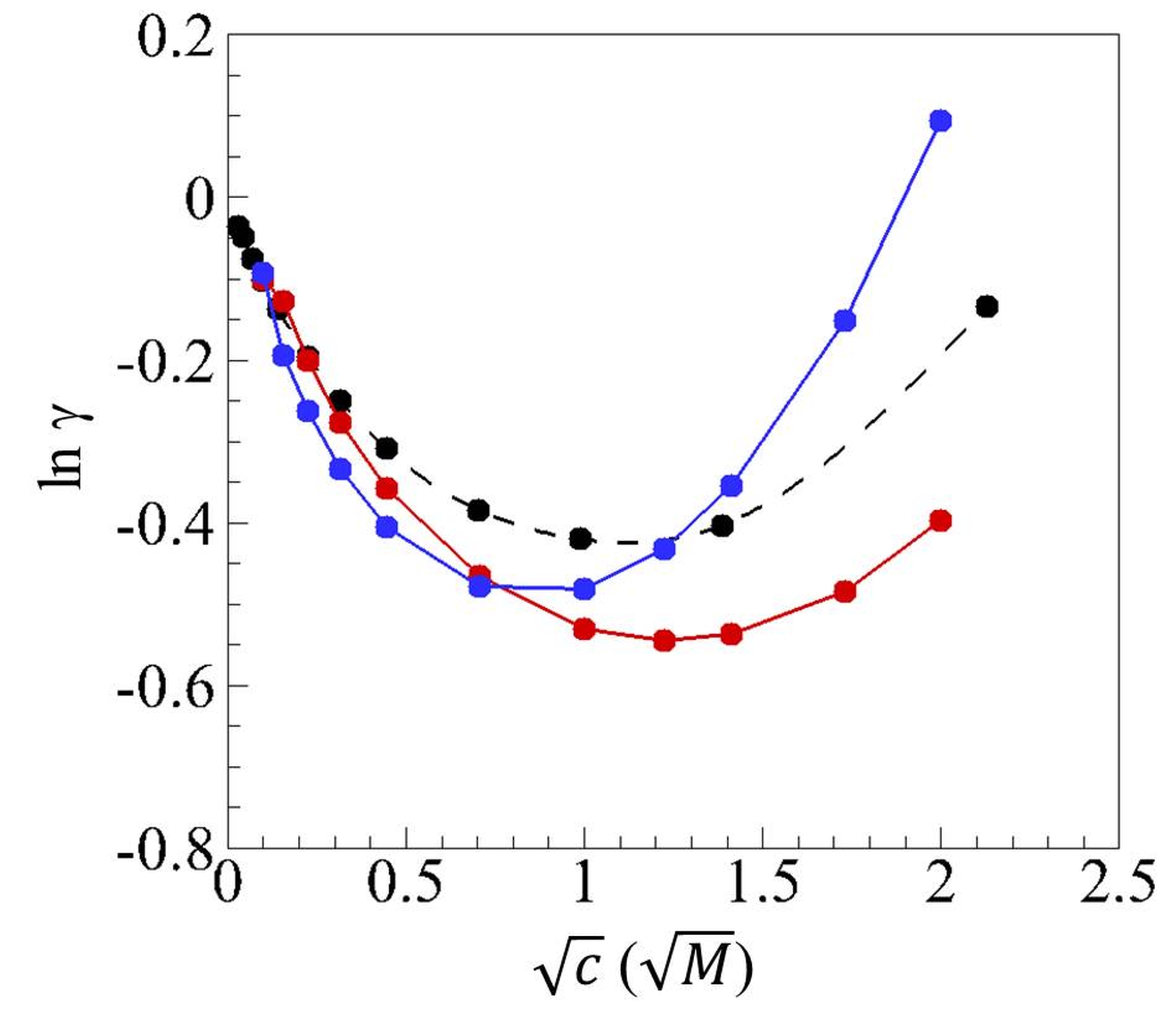}
	\centering
	\caption{Mean activity coefficient ($\gamma$) versus NaCl concentration for experiment \cite{haynes2014crc} (black) and GCMC results for the UPM (red) and SPM (blue) models.}
	\label{fig:nacl_meanactivity}
\end{figure}

This example computes the mean activity coefficients for several concentrations of aqueous sodium chloride using UPM and SPM models.
The simulation results are compared to experimental values \cite{haynes2014crc} in Figure~\ref{fig:nacl_meanactivity}.
As expected, the results diverge from experimental values as the concentration increases, with the SPM model out-performing the UPM model at higher concentrations.
These results support the conclusion \cite{vincze2010nonmonotonic} that a proper balance between solvation and Coulombic interactions is necessary for accurate activity predictions; e.g., by adding attractive ion-solvent and solvent-solvent interactions.

The simulations were performed for $8 \times 10^6$ GCMC steps using the PID method in a cubic box of 100 \r{A} side length. 
The simulation times ranged from 7 (UPM) and 12 (SPM) hours for the lowest concentration (10 mM) to 14 (UPM) and 19 (SPM) hours for the highest concentration (4 M). The 10 mM UPM simulation, however, was found to require a larger cubic box (200 \r{A} side length) to obtain statistically converged results.

\subsection*{Example 3: Calculating mono- and di-valent ion distributions around DNA}
This example uses GIBS to compute the cylindrical radial distribution and total number of ions and water around a 25bp poly(dA):poly(dT) DNA duplex, using the UPM and $\text{SPM+LJ}_\text{IW}$ models.
Bulk electrolyte solutions selected for this example are 100 mM NaCl, 100 mM RbCl, and 100 mM SrCl$_2$.
The simulations were performed with a $L_x = L_y = 150$~\r{A} and $L_z = 180$~\r{A} domain using duplex geometries from previous molecular dynamics simulations \cite{tolokh2014double}.

GCMC simulations were first performed to compute the excess chemical potential of ions and water in the bulk homogeneous system at the target concentrations, as described in the examples above.
The values computed from the two models are shown in Table S1 (Supplementary material), and were used in the DNA simulations.

\begin{figure}
	\centering
	\includegraphics{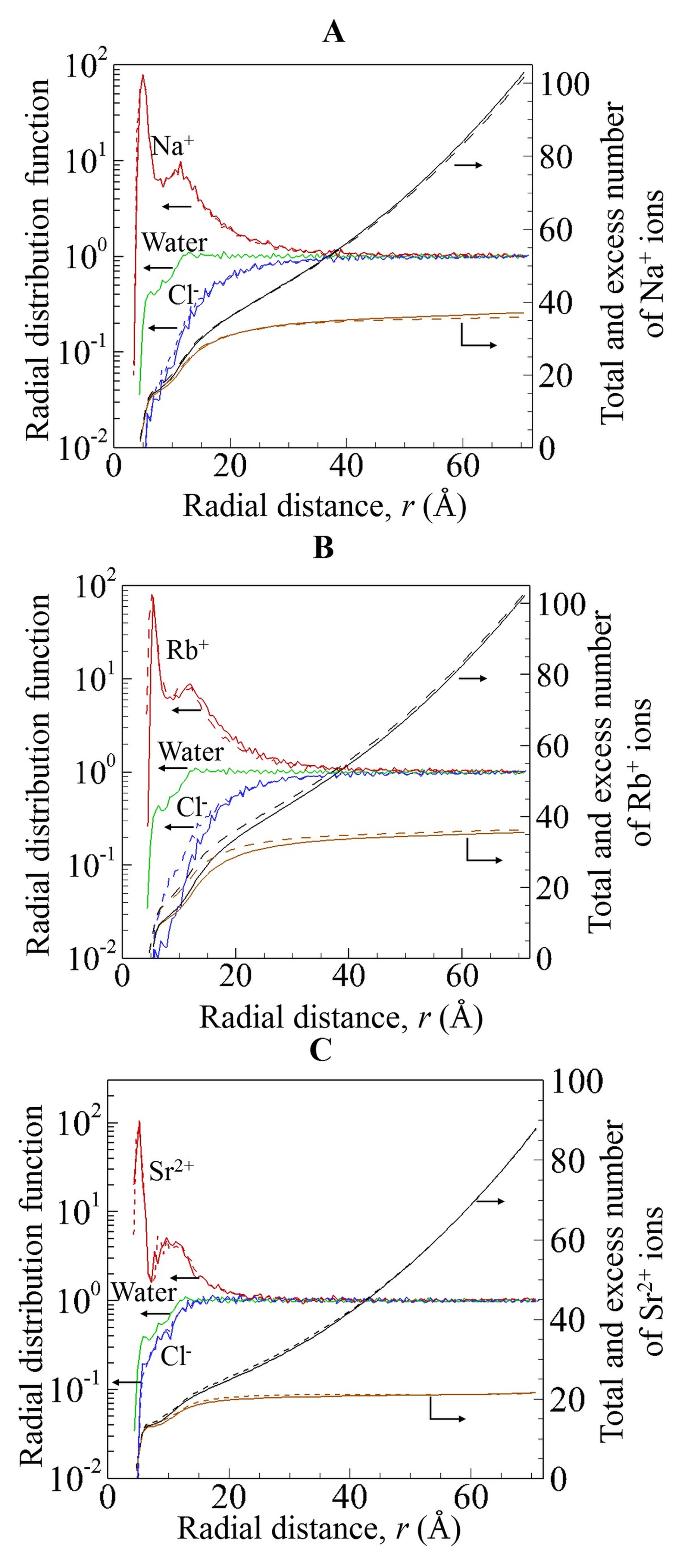}
	\caption{Cylindrical radial distribution functions of cations (red), anions (blue), and waters (green) around a poly(dA):poly(dT)duplex.
	Also shown are the total (black) and excess (brown) number of cations.
	These plots show results for (A) 100 mM NaCl, (B) 100 mM RbCl, and (C) 100 mM SrCl$_2$ solutions.
	Solid and dashed lines represent results from the UPM and SPM+LJ$_\text{IW}$ models, respectively.
	Left arrows indicate the ordinate axis of the radial distribution function. Right arrows indicate the ordinate axis of the total and excess number of cations.}
	\label{fig:polyAT25_NaRbSr}
\end{figure}

Each solute simulation was run for $8\times10^6$ GCMC steps with $\text{SPM+LJ}_\text{IW}$ and for $10^8$ GCMC steps with UPM, starting with a random configuration (state) of particles species in the simulation box.
The number of GCMC steps were determined based on how many steps it takes for the number of each particle species to converge to a stationary distribution. 
Six additional simulations of $10^7$ steps each for the $\text{SPM+LJ}_\text{IW}$ and four new simulations of $10^8$ steps each for the UPM were performed starting with different equilibrium configurations of the system obtained from the last 30\% of the first simulation. 
The UPM simulations take about 24 hours (for NaCl) and 17 hours (for RbCl and $\text{SrCl}_2$) to complete, and the $\text{SPM+LJ}_\text{IW}$ simulations take about 50 hours (for NaCl and RbCl) and 43 hours (for  $\text{SrCl}_2$) to complete.

To compute the cylindrical radial distribution function (RDF) of each particle species around DNA, the region surrounding the NA duplex was divided into cylindrical shells of width 0.5 \r{A} and length equal to 80\% of the NA duplex length (to minimize Coulombic end effects).
The RDFs from the  production runs (six for $\text{SPM+LJ}_\text{IW}$ and four for UPM) were averaged to obtain the final RDFs shown in Figure~\ref{fig:polyAT25_NaRbSr} for NaCl, RbCl, and SrCl$_2$ solutions.
The radial distance at which cation number density approaches the bulk value \cite{kirmizialtin2012rna} changes significantly with changing solvent-ion interaction models, as shown in Table~\ref{tab:excess_number}.
The excess numbers of bound Na$^+$ and Sr$^{2+}$ ions (Table~\ref{tab:excess_number}) match the ASAXS experimental values \cite{pabit2010counting} of $34 \pm 3$ and $19 \pm 2$, respectively.

\begin{table}
	\centering
	\caption{\label{tab:excess_number} Excess numbers of cations and R$_\text{bulk}$, the radial distance at which the cation number density approaches bulk, based on UPM and SPM+LJ$_{\text{IW}}$ models.}
	\begin{tabular}{|>{\centering\arraybackslash}m{2.0 cm}|>{\centering\arraybackslash}m{1.5 cm}|>{\centering\arraybackslash}m{1.0 cm}|>{\centering\arraybackslash}m{1.5 cm}|}
		\hline
		Bulk electrolyte & Model & R$_\text{bulk}$(\r{A}) & Excess number of cations  \\
		\hline
		\multirow{2}{*}{\parbox{3cm}{100 mM NaCl}} & UPM & 33.5 & 34.0 \\
		\cline{2-4}
		& SPM+LJ$_{\text{IW}}$ & 34.0 & 34.4 \\
		\hline
		\multirow{2}{*}{\parbox{3cm}{100 mM RbCl}} & UPM & 32.2 & 34.0  \\
		\cline{2-4}
		& SPM+LJ$_{\text{IW}}$ & 36.5 & 33.3 \\
		\hline
		\multirow{2}{*}{\parbox{3cm}{100 mM SrCl$_2$}} & UPM & 19.8 & 20.6 \\
		\cline{2-4}
		& SPM+LJ$_{\text{IW}}$ & 22.3 & 20.0 \\
		\hline
	\end{tabular}
\end{table}

\subsection*{Example 4: Trivalent ion distributions around DNA, RNA, and DNA-RNA hybrid duplexes}
This example uses GIBS with the UPM solvent model to determine the radial distribution of CoHex$^{3+}$ from the helical axis of four NA duplexes studied previously \cite{tolokh2014double}: B-DNA (poly(dA):poly(dT)), mixed sequence DNA, DNA:RNA hybrid, and A-RNA.
Double-stranded DNA and RNA carry the same negative charge.
The addition of small amounts of multivalent ions to DNA solutions results in inter-DNA attraction and eventual condensation; however, the same effect is not observed in RNA solutions.
Based on experimental observations and molecular dynamics simulations \cite{tolokh2014double}, it appears that this difference between DNA and RNA is due to the spatial variation in the binding of multivalent ions (CoHex$^{3+}$) to nucleic acids.

GCMC simulations were performed and the distributions were computed as described in example 3, and each production run took about 16 hours to complete.
Cylindrical radial distribution functions of CoHex$^{3+}$ (Figure S1), indicate that CoHex${^3+}$ binds differently to the four NA duplexes. Table~\ref{tab:cohex_binding} shows the average number of externally (12-16 \r{A}), internally (7-12 \r{A}), and deeply buried (0-7 \r{A}) CoHex$^{3+}$ ions around the NA helical axis. Tolokh \textit{et al} \cite{tolokh2014double} noted a positive correlation between the number of externally bound CoHex$^{3+}$ ions and the experimental duplex condensation propensity; this correlation is also observed in the GIBS results with a much simpler simulation. 

\begin{table*}
	\centering
	\caption{\label{tab:cohex_binding} Average number of CoHex$^{3+}$ ions in each binding shell and duplex neutralization, based on the UPM model.} \mbox{} \\
	\begin{tabular}{>{\centering\arraybackslash}m{3.5 cm}|>{\centering\arraybackslash}m{1.5 cm}|>{\centering\arraybackslash}m{2.0 cm}|>{\centering\arraybackslash}m{2.0 cm}|>{\centering\arraybackslash}m{2.0 cm}}
		\hline
		& DNA (dA:dT) & mixed DNA & hybrid DNA:RNA & RNA \\
		\hline
		Deeply buried (0-7\r{A})& 0.23 & 1.0 &	5.6 & 9.8 \\
		Internal shell (7-12\r{A})& 7.9 & 7.2 &	3.8 & 8.6 \\
		External shell (12-16\r{A})& 6.2 & 5.9 & 5.1 & 0.1 \\
		Duplex neutralization & 89\% & 88\% & 91\% & 120\% \\
		\hline
	\end{tabular}
\end{table*}

\section*{Conclusions}
GIBS is a new GCMC software package written in C++ and with efficient algorithms to characterize the ionic and solvent environment of biomolecules in solution.
GIBS can be used as a platform for evaluating new coarse-grained models for predicting the thermodynamic properties of ionic solutions.
The current version is freely available to the community for download at \url{https://github.com/Electrostatics/GIBS}.
We have included examples of using GIBS to compute the excess chemical potential of individual ions and salt mean activity coefficients in bulk electrolyte solutions as well as ion distributions around fixed all-atom models of nucleic acids.

Currently in GIBS, \textit{ion-ion} interactions are modeled using Coulomb, hard-sphere, and Lennard-Jones potentials; \textit{solvent-ion} interactions are modeled using hard-sphere and L-J potentials while \textit{solvent-solvent} interactions are modeled using hard-sphere repulsion potentials. 
GIBS also provides a framework for using look-up tables of ion-ion PMFs derived from explicit water MD simulations \cite{luksic2014using}, thus avoiding the need to incorporate solvent effects explicitly in the GIBS simulations and enabling faster and more accurate computations for bimolecular systems.
The resulting models may capture the dielectric environment of solvent-shared and solvent-separated ion pairs but will not account for the effect of ion concentration on the solvent structure around ions in the ion pairs or near solutes.
Therefore, future work can be focused on developing GIBS to efficiently use and to evaluate tabulated PMFs and ion solvation models that incorporate ion concentration effects.
For example, GIBS can be initially applied to develop and evaluate models for predicting the non-monotonic salt concentration dependence of mean activity coefficients in monovalent, divalent, and trivalent salt solutions \cite{vincze2010nonmonotonic}.

In GIBS solute-ion simulations, the biomolecular solute is represented as a rigid molecule using all-atom models. Coarse-grained solute models (e.g., cylindrical DNA models, spherical macroion) can also be implemented.
Since the molecule is rigid, solute conformational changes due to ion binding or ionization states (as in proteins) are not captured in the simulations.
Performing solute conformational sampling with explicit ion simulations is computationally expensive, when compared to methods using PB continuum electrostatics, as in the Multi-Conformation Continuum Electrostatics (MCCE) software program \cite{georgescu2002combining,alexov1997incorporating}.

GIBS is designed to use coarse-grained solvent and ion models to improve performance.
One coarse-grained water model implemented in GIBS is the monoatomic water (mW) model \cite{molinero2008water}.
The mW model fits well with SPM's hard sphere representation and can capture the thermodynamics and the tetrahedral structure of water without using long-ranged electrostatics to describe the short-ranged hydrogen bonding interactions.
In the mW model, each water molecule is represented as an atom with tetrahedral interactions described using the Stillinger-Weber (SW) potential \cite{stillinger1985computer}.
The re-parameterized form of the short-range SW potential has been found to faithfully reproduce both the structure and thermodynamic properties of water \cite{molinero2008water}. 
GIBS can be extended to include coarse-grained (hard-sphere chain) models of flexible polyions, such as spermine and spermidine. 

\section*{SUPPLEMENTARY MATERIAL}

\ack{An online supplement to this article can be found by visiting BJ Online at http://www.biophysj.org/biophysj/supplemental/XYZ}\vspace*{-3pt}

\section*{AUTHOR CONTRIBUTIONS}
\ack{DGT developed the GIBS software, performed the simulations, compiled the results, and drafted the manuscript. Both DGT and NAB designed the research, analyzed the results, and wrote the final manuscript.}
\section*{ACKNOWLEDGMENTS}

\ack{This work was supported by NIH grants GM076121 and GM099450.
The authors thank Alexey Onufriev and Igor Tolokh for providing PDB structure files of the nucleic acid duplexes as well as Lois Pollack, Suzette Pabit, Maria Sushko, and Marcel Baer for helpful discussions.
The authors also thank Chenguang Li for helping to implement the spline interpolation methods in GIBS. Pacific Northwest National Laboratory is operated for DOE by Battelle Memorial Institute under contract DE-AC05-76RL01830.}\vspace*{6pt}
\bibliography{references}

\end{document}